\documentstyle[twocolumn,prl,aps,psfig]{revtex}
\begin{document}
\draft
\narrowtext
\title{\Large\bf Novel diffusion mechanism on the GaAs\,(001) surface: 
the role of adatom--dimer interaction   }
\author{Alexander Kley, Paolo Ruggerone, and Matthias Scheffler}
\address{
Fritz-Haber-Institut der Max-Planck-Gesellschaft,\\ Faradayweg 4-6,
D-14195 Berlin (Dahlem), Germany}
\date{\today}

\maketitle

\begin{abstract}
  Employing first principles total energy calculations we have studied the
  behavior of Ga and Al adatoms on the GaAs(001)--$\beta2$ surface.  The
  adsorption site and two relevant diffusion channels are identified.  The
  channels are characterized by different adatom--surface dimer interaction.
  Both affect in a novel way the adatom migration: in one channel the
  diffusing adatom jumps across the surface dimers and leaves the dimer bonds
  intact, in the other one the surface dimer bonds are broken. The two
  channels are taken into account to derive effective adatom diffusion
  barriers.  From the diffusion barriers we conclude a strong diffusion
  anisotropy for both Al and Ga adatoms with the direction of fastest
  diffusion parallel to the surface dimers.  In agreement with experimental
  observations we find higher diffusion barriers for Al than for Ga.
\end{abstract} 

Growth techniques, such as molecular beam epitaxy, operate under conditions
far away from thermodynamic equilibrium.  Particularly for growth at low
temperatures or for structures with length scales smaller than the adatom
diffusion length, features driven by the growth kinetics have been
observed.~\cite{koshiba:94} In the GaAs/AlAs heteroepitaxy the differences
between the growth kinetics of AlAs and GaAs have been utilized to create low
dimensional structures.~\cite{shen:94} Recently Kapon {\it et
  al.}\cite{kapon:89} successfully fabricated quantum wire heterostructure in
which stimulated emission has been observed.

Despite these successes the underlying microscopic processes such as
adsorption, surface diffusion, desorption, and nucleation are poorly
understood. A key mechanism in growth is cation surface diffusion which is
considered to be a rate limiting process.~\cite{kawabe:89,horikoshi:86}
Experimentally the surface diffusion is difficult to access.  The deduced
migration barriers\cite{nishinaga:88,neave:85,shitara:93,ohta:89,hove:87} for
Ga adatoms span a range between 1.1 and 4.0\,eV.  Even the anisotropy of
surface diffusion on the (001) surface is controversially debated: Shitara
{\it et al.}~\cite{shitara:92b} speculate that fastest diffusion is along the
[110] direction; in contrast, Kawabe {\it et al.}~\cite{kawabe:89} propose the
[$\overline{\rm 1}$10] direction.  We have therefore performed
first-principles total-energy calculations which constitute a powerful tool to
study surface diffusion.

Adatoms on solid surfaces occupy well defined binding sites. The migration of
the chemisorbed adatoms can be described as a hopping between these sites.
The activation energies for the individual hops are determined by the energy
differences between the binding and the transition sites.  These positions can
be identified as the minima and saddle points of the potential energy surface
over the configurational space spanned by the the coordinates of the adatom
and the substrate atoms.  In order to find all minima and transition sites a
mapping of the entire configurational space is in principle required. However,
this is computationally neither possible nor useful.  For the study of surface
diffusion the mapping is commonly restricted to a subspace given by the
lateral coordinates of the
adsorbate.~\cite{brocks:91,yu:94,kaxiras:96,shiraishi:96,gilmer:92} This
mapping gives the potential energy surface (PES) $E(x,y)$ for a given lateral
position $(x,y)$ of the adatom where all substrate atoms and the $z$
coordinate of the adatom are fully relaxed. This implies that all minima and
saddle points relevant for the description of surface diffusion can be found
on one single PES.  However, in this Letter the results of our first
principles study on the cation diffusivity show that this assumption fails for
the migration on the GaAs(001) surface.

\begin{figure}
\psfig{figure=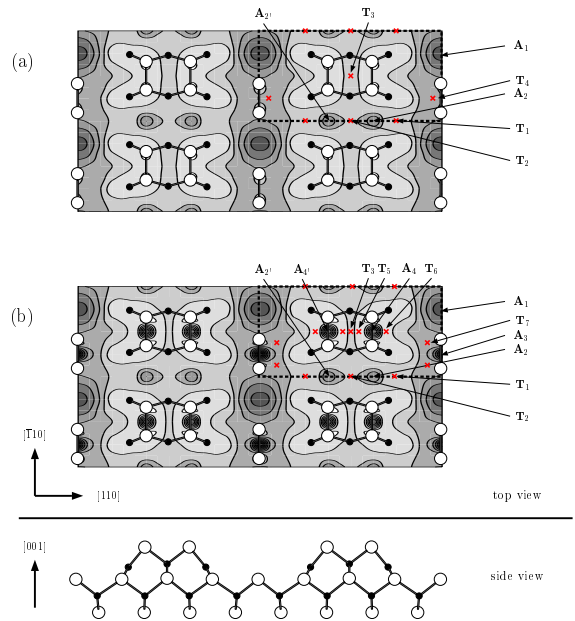,width=9cm}
\caption{
  Total energy surfaces for a Ga adatom on the GaAs(001)--(2$\times$4)
  $\beta2$ surface. (a)~PES obtained when the adatom is relaxed from 3\,$\rm
  \AA$ above the surface\ (b)~PES obtained when the adatom is relaxed from
  0.5\,$\rm \AA$ above the surface with the surfac\ e dimers initially broken.
  The dashed box shows the unit cell.  The contour--line spacing is 0.2\,eV.
  The atomic positions of the clean surface are indicated for atoms of the
  upper two layers and for the As-dimers in the third layer
  (As:~empty~circles, Ga:~filled~circles).  }\label{fig:1}
\end{figure}

In this paper we focus on the GaAs(001)--(2$\times$4) surface in the $\beta2$
phase. This structure has been shown to be stable at the
equilibrium~\cite{northrup:94} as well as to be the dominating surface
structure over a wide range of growth
conditions.~\cite{daeweritz:90,hashizume:94} As shown in Fig.\ \ref{fig:1},
each (2$\times$4) unit cell consists of two As dimers and two missing dimers
in the topmost layer, a missing Ga pair in the second layer and a As dimer in
the third layer.

Our calculations of the PESs employ density functional theory in the
generalized gradient approximation (GGA).~\cite{perdew:92} The electron-ion
interaction is described by fully separable, norm-conserving
pseudopotentials.~\cite{bachelet:82,bylander:90} The adatom-substrate system
is modeled by a supercell geometry with a (4$\times$4) periodicity parallel to
the surface. This supercell has been tested to be sufficiently large to have a
negligible adatom-adatom interaction. Perpendicular to the surface our
supercell contains a vacuum of 6 layers GaAs and an additional layer of pseudo
H-atoms (Z=0.75) to saturate the bonds of the lower
surface.~\cite{shiraishi:90} To prevent a dipole-dipole interaction between
the inequivalent upper and lower surface of our slab we use a dipole
correction.~\cite{neugebauer:92} The wave functions are expanded in a
plane-wave basis with a cutoff energy of 10\,Ry. The {\bf k}--space
integration was performed with a special {\bf k}-point set, with a density
equivalent to 64 {\bf k}-points in the Brillouin zone of the (1$\times$1)
surface cell.

The PES has been mapped on an equidistant grid with a spacing of 1\,$\rm \AA$
along the [110] and the [$\overline{\rm 1}$10] direction.  At each position
the upper four substrate layers and the adatom height were fully relaxed,
starting from the positions of the clean surface for the substrate and an
adatom height of 3$\rm\,\AA$ above the surface.  In a second step we
interpolated the PES from the energy values on the mesh.  Subsequently we
repeated our mapping around the interpolated saddle points using a finer mesh.
The exact positions of the local minima were obtained by putting the adatom on
the interpolated minima and relaxing {\em all} atomic positions without
constraints.

The resulting PES for a Ga adatom is shown in Fig.\ \ref{fig:1}(a).  In each
(2$\times$4) unit cell we find the binding sites: $\bf A_{\rm 1}$ and the two
equivalent positions $\bf A_{\rm 2}$ and $\bf A_{\rm 2'}$.  All three binding
sites are on long bridge positions between surface As--dimer.  This agrees
with previous first principles calculations~\cite{shiraishi:96} for the
metastable (2$\times$4)--$\beta$ structure (assuming near-equilibrium growth
conditions~\cite{tersoff97}, at $T = 900$ K, a typical growth temperature, the
surface area covered with the $\beta2$ phase and the area covered by the
$\beta$ phase differ by more than two orders of magnitude).  At the three
binding sites our calculations yield a strong inward relaxation of the adatom
which resides almost in the center between the neighboring As atoms.  The
substrate atoms display only minor deviations ($< 0.2\,\rm \AA$) from their
clean surface positions; in particular the surface dimer bonds remain intact.

From PES shown in Fig.\ \ref{fig:1}(a) migration on GaAs(001) can be described
as a sequence of random hops on a periodic (2$\times$4) lattice with the three
binding sites.  The effective diffusion barrier is determined by applying the
formalism of continuous time random walk (CTRW)~\cite{haus:87} that yields the
diffusion tensor as a function of the individual hopping rates between
neighboring binding sites.  Within transition state theory~\cite{vineyard:57}
the individual hopping rate from a site ${\bf A_{\rm i}}$ to a site ${\bf
  A_{\rm j}}$ via a transition state $\bf T_{\rm k}$ can be written as
\begin{equation}
  \Gamma_{\rm ij}=\Gamma_{\rm ij}^{\rm 0}\;\exp[(E({\rm \bf T_{\rm k}})-E({\rm
    \bf A_{\rm i}}))/k_{\rm B} T]
\end{equation} 
where $E({\rm \bf A_{\rm i}})$ and $E({\rm \bf T_{\rm k}})$ are
the energies given in Tab.\ \ref{tab:binding}.  $\Gamma_{\rm ij}^{\rm 0}$ is a
temperature independent prefactor and assumed to be roughly the same for all
individual hops.  This assumption leads to only minor errors as long as solely
the effective diffusion barrier are
discussed.~\cite{brocks:91,yu:94,shiraishi:96} Applying the CTRW
formalism~\cite{kley:97} and considering the size of the individual activation
energies we find for the diffusion constants along the [110] and
[$\overline{\rm 1}$10] directions
\begin{eqnarray}
  D_{\rm [110]} = & \frac{\Gamma_{\rm 12} \Gamma_{\rm 21} ( 2 \Gamma_{\rm 22}
    + \Gamma_{\rm 22'})} {( 2 \Gamma_{\rm 12} + \Gamma_{\rm 21})(4 \Gamma_{\rm
      22} + \Gamma_{\rm 21}+2 \Gamma_{\rm 22'})}\;8 a_{\rm 0}^{\rm 2} &
  \approx \frac{\Gamma_{\rm 12}}{2}\;8 a_{\rm 0}^{\rm 2} \label{eq:some_A_s}\\ 
  D_{\rm [\overline{\rm 1}10]} =& \frac{4 \Gamma_{\rm 12} \Gamma_{\rm 22} +
    \Gamma_{\rm 11} \Gamma_{\rm 21}}{2 \Gamma_{\rm 12} + \Gamma_{\rm 21}} \; 2
  a_0^{\rm 2} & \approx \Gamma_{\rm 11}\; 2 a_0^{\rm 2}\label{eq:some_A_p}
\end{eqnarray} 
$a_{\rm 0}$ is the lattice constant of GaAs.  From the
approximated solutions we get as effective diffusion barriers $\Delta E_{\rm
  [110]}= E({\rm \bf T_{\rm 1}})-E({\rm \bf A_{\rm 1}})$ and $\Delta E_{\rm
  [\overline{\rm 1}10]}= E({\rm \bf T_{\rm 4}})-E({\rm \bf A_{\rm 1}})$, i.e.
$\Delta E_{\rm [110]}= 0.8$\,eV and $\Delta E_{\rm [\overline{\rm 1}10]}=
0.6$\,eV.

In the above discussion we have implicitly assumed that all relevant minima in
the configuration space could be found by relaxing the adatom from a position
in the vacuum and the substrate from its ideal position in the clean case.
This is true if all other minima of the total energy are either energetically
much higher or separated from the first minima by a large barrier.  The
assumption should be correct when the surface exhibits mainly bulk-like bonds,
as it is the case for most metal surfaces.  For reconstructed semiconductor
surfaces, however, surface bonds exist which differ significantly from the
bulk bonds, as for example anion--anion dimers. On those surfaces stable
adsorption sites for a cation could be realized by breaking these weak surface
bonds and forming bulk-like cation-anion bonds instead.

\begin{figure}
\psfig{figure=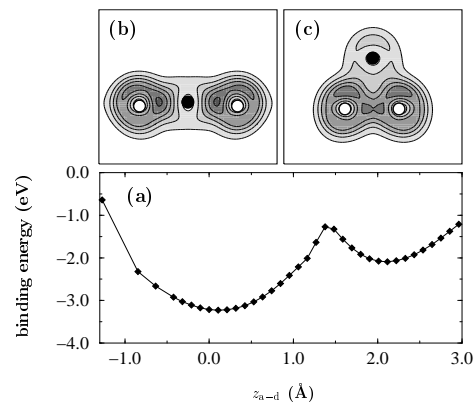,width=8cm}
\caption{
  (a) Binding energy of an Ga adatom as function of its height $z_{\rm a-d}$
  above the center of an As-dimer. (b) and (c) charge density cross section
  through the adato\ m and the dimer atoms at the two minima of the binding
  energy.  }\label{fig:2}
\end{figure}

Since in the PES shown in Fig.~\ref{fig:1}(a) the adatom does not break any
surface As-dimer, we investigate in more details the adatom--dimer
interaction. Fig.\ \ref{fig:2}(a) shows the binding energy of the adatom as a
function of its height $z_{\em a-d}$ on the dimer center with its lateral
coordinates ($x,y$) fixed above the dimer center [site ${\bf A_{\rm 3}}$ in
Fig.\ \ref{fig:1}(b)] and the substrate fully relaxed.  Approaching the
surface from the vacuum we find a first local minimum at $z_{\em a-d} \approx
2\,\rm \AA$ which corresponds to a saddle point in the PES shown in Fig.\ 
\ref{fig:1}(a). Coming closer to the surface there is a barrier at $z_{\em
  a-d} \approx 1.2\,\rm \AA$ and another much deeper minimum at $z_{\em a-d}
\approx 0.1\,\rm \AA$.  Such a double potential well was already predicted for
Si on Si\,(001) by Roland and Gilmer.~\cite{gilmer:92} The origin of the
barrier and the presence of the second minimum can be understood by analyzing
the bonding situation of the adatom to the dimer. At the first minimum the
adatom interacts with the completely filled dangling bonds of the dimer
[Fig.~\ref{fig:2}(c)], at the second one it breaks the dimer bond and forms
directional bonds with the adjacent anions [Fig.~\ref{fig:2}(b)]. The barrier
between the minima is due to the energy cost required to break the As--dimer.
The binding energy at the second minimum is higher than that at the previously
found binding sites $\bf A_{\rm 1}$, $\bf A_{\rm 2}$ and $\bf A_{\rm 2'}$.
This shows that the adsorption site which is defined as the site with the
highest binding energy cannot be on the PES displayed in Fig.\ \ref{fig:1}(a).

To find the real adsorption site and to include the influence of the dimer
breaking on the diffusion properties we map the PES a second time. In contrast
to the mapping of the first PES the adatom height $z$ is now relaxed starting
from a position 0.5\,$\rm \AA$ above the surface and the surface As atoms are
relaxed from a initial position where the As--As dimer bond length is
3.5\,$\rm \AA$, i.e.  the directional dimer bonds are broken.  Fig.\ 
\ref{fig:1}(b) displays the resulting PES. Figs.\ \ref{fig:1}(a) and (b) show
that the two PESs are similar over a wide area but they differ significantly
around the dimers.  In addition to the binding sites already present on the
first PES we find three new binding sites in each unit cell: ${\rm \bf A_{\rm
    3}}$ and the two equivalent positions ${\rm \bf A_{\rm 4}}$ and ${\rm \bf
  A_{\rm 4'}}$.  The comparison of the binding energies of the whole set of
binding sites yields as adsorption site ${\bf A_{\rm 3}}$, a short bridge site
between the As atoms of one dimer. 

At ${\rm \bf A_{\rm 3}}$, ${\rm \bf A_{\rm 4}}$ and ${\rm \bf A_{\rm 4'}}$ the
adatom breaks the dimer, and charge accumulations between the adatom and the
adjacent anions confirm the formation of directional bonds.  The angle between
the two adatom-As bonds is about 175$^{\circ}$, i.e. the adatom is almost
incorporated into the As top layer.  This is accompanied by a strong
horizontal relaxation of the As atoms bonded to the adatom.  They relax along
the [$\overline{\rm 1}$10] direction, and the As--As distance increases from
$\approx$ 2.5\,\AA~(the As--dimer bond length) to 4.4\,\AA~which is close to
the As-As distance in bulk GaAs (4.0\,\AA). The relaxation keeps the nearest
neighbor bond length of the substrate atoms nearly unchanged and results in
bulk like adatom--As bond lengths.  Such a relaxation has been observed
experimentally for an Al terminated (2$\times$1) reconstructed GaAs (001)
surface.~\cite{bachrach:81} The driving mechanism for the flat incorporation
of the cation is analogous to that which leads to the well known inward
relaxation of the surface Ga atoms on GaAs(110).  The two--fold coordinated
adatom with only partially occupied dangling bonds rehybridizes to a planar
$sp$ or $sp^2$ bonding configuration.  At the same time the three--fold
coordinated As atoms bonded to the adatom gain energy from rehybridization to
a $p^3$ bonding configuration.

To find the exact binding energies at the transition sites ${\rm \bf T_{\rm
    5}}$, ${\rm \bf T_{\rm 6}}$ and ${\rm \bf T_{\rm 7}}$ between the short
bridge and the long bridge binding sites the PES has been mapped around the
interpolated transitions sites as a function not only of the lateral adatom
coordinates but also of the dimer atoms. Note that we found an energy barrier 
to break the dimer for every pathway of the adatom toward ${\bf{A}}_3$, 
${\bf A_{\rm 4}}$, and ${\bf A_{\rm 4'}}$.

Including the new binding sites, surface diffusion is described by a random
hopping on a periodic (2$\times$4) lattice with six sites in each unit cell:
${\bf A_{\rm 1}}$, ${\bf A_{\rm 2}}$, ${\bf A_{\rm 2'}}$, ${\bf A_{\rm 3}}$,
${\bf A_{\rm 4}}$, and ${\bf A_{\rm 4'}}$.  With the CTRW formalism we
approximate the diffusion constants (the rather long exact expression will be
published elsewhere\ \cite{kley:97}) as:
\begin{eqnarray}
  D_{\rm [110]} & \approx & \frac{ \Gamma_{\rm 12} \Gamma_{\rm 31} +
    \Gamma_{\rm 13} \Gamma_{\rm 32} } {2 \Gamma_{\rm 13}}\; 8 a_{\rm 0}^{\rm
    2} \\ D_{\rm [\overline{\rm 1}10]} & \approx & \frac{ \Gamma_{\rm 11}
    \Gamma_{\rm 31}}{ \Gamma_{\rm 13}} \; 2 a_0^{\rm 2}\label{eq:all_A_p}
\end{eqnarray}
From theses equations we find as effective diffusion barriers $\Delta E_{\rm
  [110]}= E({\rm \bf T_{\rm 1}})-E({\rm \bf A_{\rm 3}})$ and $\Delta E_{\rm
  [\overline{\rm 1}10]}= E({\rm \bf T_{\rm 4}})-E({\rm \bf A_{\rm 3}})$.
Inserting the energies given in Tab.\ \ref{tab:binding} we find for the Ga
adatom diffusion $\Delta E_{\rm [110]}= 1.5$\,eV and $\Delta E_{\rm
  [\overline{\rm 1}10]}= 1.2$\,eV.  From these values a significant diffusion
anisotropy with fastest diffusion parallel to the surface dimers can be
concluded.

The inclusion of the binding sites of the second PES changes the diffusion
barriers dramatically. At first glance, taking into account only the second
PES seems to be sufficient for the calculation of the diffusion barriers.
However, from Eq. (\ref{eq:all_A_p}) it is clear that the description of the
diffusion parallel to the surface dimers demands hopping rates from both PESs.
If the contributions from one PES are neglected [as in the Eqs.
(\ref{eq:some_A_s}) and (\ref{eq:some_A_p})], wrong barriers are obtained. The
reason for involving both PESs is that the energy barrier to hop from a site
${\bf A_{\rm 1}}$ to a site ${\bf A_{\rm 3}}$ (which requires the breaking of
a dimer) is only slightly higher than the one to migrate from the site ${\bf
  A_{\rm 1}}$ to a neighboring ${\bf A_{\rm 1}}$ site. Thus, once an adatom
occupies a weakly bound ${\bf A_{\rm 1}}$ site it performs a number of hops to
neighboring ${\bf A_{\rm 1}}$ sites (like surfing on the first PES) before
being trapped again in a strongly bound ${\bf A_{\rm 3}}$ site.

Our calculated diffusion barriers for a Ga adatom are at the lower limit of the
experimentally deduced diffusion barriers (1.1 -- 4.0\,eV). This seems 
reasonable because the experimental results were derived rather indirectly
from growth experiments and therefore they are affected by adatom--adatom 
and adatom--step interactions. Further, the
adatom mobility depends on the As supply and decreases with increasing As
flux.~\cite{hata:91} Accounting for such effects should result in an effective
diffusion barrier higher than that calculated here for a single adatom.

We have performed similar calculations also for an Al adatom. The PESs are
almost identical to those of a Ga adatom, i.e.  the positions of the minima
and of the saddle points for the two cations differ by less than 0.1\,$\rm
\AA$. The main difference is a slightly stronger corrugation for Al adatoms
leading to higher diffusion barriers: $\Delta E_{\rm [110]}= 1.6$\,eV and
$\Delta E_{\rm [\overline{\rm 1}10]}= 1.3$.  The enhanced corrugation can be
understood in terms of the higher cohesive energy of AlAs ($E_{\rm AlAs}^{\rm
  coh} = 3.78$\,eV) compared to GaAs ($E_{\rm GaAs}^{\rm coh} = 3.26$\,eV)
implying that Al-As bonds are stronger than Ga-As bonds.  From the higher
barriers we conclude a lower mobility of Al adatoms which is in accordance
with several experiments.~\cite{koshiba:94,shitara:93} Our calculated
difference in the barrier height of about 0.1\,eV for Al compared to Ga
adatoms agrees rather well with the value of 0.16\,eV obtained by Shitara {\it
  et al.}~\cite{shitara:93} from Monte--Carlo simulations of RHEED
measurements of the growth of AlAs and GaAs.

In conclusion, we have shown that the adatom--surface dimer interaction is
crucial for determining the adsorption site as well as the effective diffusion
barrier.  The origin of the very stable new adsorption site is that a weak
surface bond (As--As) is replaced by almost bulk--like cation--As bonds. This
mechanism should also work on other semiconductor surfaces exhibiting surface
dimers or trimers.

Stimulating discussions with E. Pehlke and K. Shiraishi are acknowledged.  We
thank the SFB--296 for financial support.


\begin{table}[ht]
\caption{
  Binding energies (eV) of a Ga adatom at various surface sites on the
  GaAs(001)--(2$\times$4) $\beta2$ structure}\label{tab:binding}
\begin{tabular}{ccccccccccccc}
  site & ${\bf A_{\rm 1}}$ & ${\bf A_{\rm 2}}$ & ${\bf A_{\rm 3}}$ & ${\bf
    A_{\rm 4}}$ & ${\bf T_{\rm 1}}$ & ${\bf T_{\rm 2}}$ & ${\bf T_{\rm 3}}$ &
  ${\bf T_{\rm 4}}$& ${\bf T_{\rm 5}}$ & ${\bf T_{\rm 6}}$ & ${\bf T_{\rm 7}}$
  \\ Ga & -2.5 & -2.2 & -3.2 & -2.6 & -1.7 & -1.8 & -1.6 & -2.0 & -1.45 & -1.5
  & -1.9
\end{tabular}
\end{table}



\end{document}